\documentstyle[preprint,aps]{revtex}
\tighten
\begin{document}
\draft
\preprint{\vbox{Submitted to Phys.\ Rev.\ D \hfill FSU-SCRI-94-15 \\
                                       \null\hfill nucl-th/9402019}}
\title{On the Lorentz structure of the confinement potential}
\author{J. Parramore}
\address{Department of Physics and
         Supercomputer Computations Research Institute, \\
         Florida State University, Tallahassee, FL 32306}
\author{J. Piekarewicz}
\address{Supercomputer Computations Research Institute, \\
         Florida State University, Tallahassee, FL 32306}
\date{\today}
\maketitle
\begin{abstract}
We investigate the Lorentz structure of the confinement
potential through a study of the meson spectrum using
Salpeter's instantaneous approximation to the Bethe-Salpeter
equation. The equivalence between Salpeter's and a
random-phase-approximation (RPA) equation enables one
to employ the same techniques developed by Thouless, in
his study of nuclear collective excitations, to test the
stability of the solutions. The stablity analysis reveals
the existence of imaginary eigenvalues for a confining potential
that transforms as a Lorentz scalar. Moreover, we argue that
the instability persists even for very large values of the
constituent quark mass. In contrast, we find no evidence of
imaginary eigenvalues for a timelike vector potential --- even
for very small values of the constituent mass.
\end{abstract}
\pacs{PACS number(s): 11.10.St, 12.40.Qq}
\section{Introduction}

Quantum chromodynamics (QCD) is believed to be the correct
theory of the strong interactions. Still, the description of
hadrons in terms of the underlying degrees of freedom of the
theory, namely, quarks and gluons, remains a daunting task.
Although fundamental approaches such as lattice gauge theory
continue to improve, it is unrealistic to forsee lattice
calculations of the hadronic spectrum in the immediate future.
Yet, it may still be fruitful to use lattice QCD to determine
properties and parameters of a phenomenological model.
Nonrelativistic quark models have been quite successful in this regard.
For example, quarkonium, the bound-state spectrum of a quark-antiquark
system, is well reproduced by a phenomenological potential consisting
of the sum of a one-gluon exchange component and a confinement
contribution. The quantitative description of meson masses, their
static properties and decay rates count among the many successes of
the model.

Lattice simulations have provided conclusive evidence that QCD
is a confining theory. For large quark-antiquark $(q\bar{q})$
separations, lattice QCD predicts a linearly rising potential
between heavy quarks~\cite{creutz80}. However, heavy-quark systems
place the strongest constraints on the short-distance part of the
${q\bar{q}}$ interaction and, thus, remain fairly insensitive to
the details of the confinement potential. Thus, it is safe to conclude
that the nonrelativistic quark model reveals little about the nature
of quark confinement in the region where the model is expected to be
valid. The light-quark spectrum is sensitive to the long-range part of
the potential but the assumptions of the model are questionable in
this case. Many attempts have been made to incorporate relativistic
corrections into the nonrelativistic quark
model~\cite{stanley80,godfrey85}. For the most part these have
consisted of a relativistic dispersion relation together with
relativistic ($p/M$) corrections to the nonrelativistic potential.
It is essential to stress, however, that these dynamical corrections
can only be effected after the Lorentz structure of the potential has
been established. At present, very little is known about the dynamical
origin of the confinement potential and even less about its Lorentz
transformation properties. Understanding the Lorentz character of the
confinement potential --- by means of a stability analysis of the
instantaneous Bethe-Salpeter equation --- is the main focus of the
present work.

The starting point for most relativistic studies of the meson spectrum
is the covariant Bethe-Salpeter equation~\cite{salpeter51}. Unfortunately,
there are difficulties associated with using the Bethe-Salpeter equation
(BSE) for bound states. Chief among these is the appearance of a
relative-time variable due to retardation effects. Three-dimensional
reductions of the full BSE, which eliminate the relative-time variable,
abound. While most of these reductions attempt to preserve fundamental
physical principles, there is still no method of choice. Here, we work
within the framework of Salpeter's instantaneous
approximation to the Bethe-Salpeter equation~\cite{salpeter52}.

We believe that the use of an instantaneous approximation is an
appealing choice for the study of the Lorentz structure of the
confinement potential. We recognize that, because very little is
known about the dynamical origin of the confinement potential,
introducing non-instantaneous corrections into the formalism is
highly model dependent. We do not pretend to make any contribution
to this subject. Instead, we simply adopt the position that if an
ad-hoc modelling of retardation must be done, ignoring it altogether
is probably the best choice. Still, we recognize that a choice for
the Lorentz structure of the potential must be made before it can be
incorporated into a relativistic equation. We should also mention that
because the long-range part of the potential is the main focus of the
present work, the short-range, one-gluon exchange (OGE) component will
be ignored. If, instead, the goal becomes the realistic description of
the meson spectrum, one would be forced to include the OGE piece. Note
that since the OGE contribution has a clear dynamical origin, incorporating
retardation effects now becomes important and largely model independent.
In this case, the use of an instantaneous approximation might be harder
to justify.

The Lorentz structure of the confinement potential is usually
assumed to be scalar~\cite{gromes91}. Yet, several authors have
suggested that the hypothesis of scalar confinement is not correct.
Lag\"{a}e~\cite{lagae92}, and independently, Gara, Durand, and
Durand~\cite{gara89} examined several models within the framework
of the instantaneous Bethe-Salpeter equation and concluded that
the confining potential was not a scalar. Archvadze, Chachkhunashvili,
and Kopaleishvili examined the light-mass spectrum and concluded that
no stable solutions exist for scalar confinement~\cite{arch93}.
More recently, M\"unz, Resag, Metsch, and Petry found that reliable
solutions of Salpeter's equation for scalar confinement exist only
for large quark masses and weak confinement strength~\cite{munz93}.

In this paper we present a stability analysis of Salpeter's equation
in the pseudoscalar channel assuming, both, scalar and timelike Lorentz
structures for the confinement potential. The stability analysis stems
from the recognition that Salpeter's equation is identical in structure
to a random-phase-approximation (RPA) equation familiar from the study
of nuclear collective excitations. This stability analysis reveals the
existence of imaginary eigenvalues for the case of scalar confinement.
Moreover, an analytic study suggests that the instability should
persist even in the case of a very large constituent mass. In contrast,
no instability is observed for timelike confinement.

We have organized the paper as follows. In Sec.~\ref{secformal}
we review some general properties of the instantaneous approximation
to the Bethe-Salpeter equation. Special emphasis is placed on
the RPA structure of the equation. It is in this context that
we make contact with previous stability analyses of the nuclear
many-body problem. In Sec.~\ref{secresults} we discuss our method
of solution and present results for the mass spectrum of pseudoscalar
states for a variety of quark masses. Finally, Sec.~\ref{secconcl}
contains our conclusions and directions for future work.

\section{Formalism}
\label{secformal}
Here, we briefly outline a derivation of Salpeter's equation,
using Greens's function methods, for a fermion-antifermion pair
of equal mass (the extension to different masses is straightforward).
In addition, we show the equivalence of Salpeter's equation to an RPA
equation.
\subsection{The Salpeter equation}
\label{secsalp}
The starting point for our derivation is
the four-point Green's function, or two-body propagator,
defined by
\begin{equation}
  iG_{\alpha\beta ; \lambda\sigma}(x_1,x_2;y_1,y_2) \equiv
  \langle {\Psi_0} |
  T\left [ \psi_{\alpha}(x_1) \psi_{\beta}(x_2)
           \bar{\psi}_{\sigma}(y_2) \bar{\psi}_{\lambda}(y_1)
   \right ]
  | {\Psi_0} \rangle \;,
\label{prop}
\end{equation}
where $\psi_{\alpha}(x)$ are fermion fields in the Heisenberg
representation, $\alpha , \beta , \lambda , \sigma$ are Dirac
spinor indices, and $\Psi_0$ represents the exact vacuum wave
function. The four point function describes the propagation of
two fermions $(FF)$, or a fermion-antifermion ($F\bar{F}$) pair,
and contains all scattering and bound-state information. Restricting
the relative time variables in the Green's function, enables one
to select out, either, $FF$ or $F\bar{F}$ modes from the propagator.

The two-body propagator, and ultimately the bound-state spectrum,
will be generated as a solution to the instantaneous Bethe-Salpeter
equation in the ladder approximation. This is the central assumption
of the model. Thus, in this approximation the irreducible kernel is
given by
\begin{equation}
  V(x_{1},x_{2}) \equiv V({\bf x_{1}},{\bf x_{2}}) \;.
 \label{vinst}
\end{equation}
In Salpeter's instantaneous approximation there are two
remaining time variables (and, thus, one remaining relative
time variable). By selecting them according to
\begin{eqnarray}
  x_{1}^{0} &=& y_{2}^{0} \equiv t  \;, \\
  x_{2}^{0} &=& y_{1}^{0} \equiv t' \;,
\end{eqnarray}
one isolates the propagation of the $F\bar{F}$, as opposed to the
$FF$, mode. In this limit, Salpeter's equation for the two-body
propagator becomes,
\begin{eqnarray}
   G_{\alpha\beta;\lambda\sigma}
   ({\bf x}_1,{\bf x}_2;{\bf y}_1,{\bf y}_2;\omega) =
   G^{(0)}_{\alpha\beta;\lambda\sigma}
   ({\bf x}_1,{\bf x}_2;{\bf y}_1,{\bf y}_2;\omega) \nonumber \\ +
   \int d{\bf z}_1 \; d{\bf z}_2
   G^{(0)}_{\alpha\eta^{\prime};\xi\sigma}
   ({\bf x}_1,{\bf z}_2;{\bf z}_1,{\bf y}_2;\omega)
   V_{\xi\eta ;\xi^{\prime}\eta^{\prime}}({\bf z}_1,{\bf z}_2)
   G_{\xi^{\prime}\beta;\lambda\eta}
   ({\bf z}_1,{\bf x}_2;{\bf y}_1,{\bf z}_2;\omega) \;,
\label{salpint}
\end{eqnarray}
where $G^{(0)}$ is the free two-body propagator and $\omega$ is
the energy variable conjugate to $(t-t')$. A study of the
analytic structure of $G$, by means of a Lehmann representation,
reveals that the propagator contains singularities at the position
of the bound-state poles. Salpeter's equation can then be obtained
by picking up the residue at the bound-state pole
\begin{equation}
  \chi^{E}_{\alpha\sigma}({\bf x}_1,{\bf y}_2) =
  \int d{\bf z}_1 \; d{\bf z}_2
   G^{(0)}_{\alpha\eta^{\prime};\xi\sigma}
   ({\bf x}_1,{\bf z}_2;{\bf z}_1,{\bf y}_2;E)
   V_{\xi\eta ;\xi^{\prime}\eta^{\prime}}({\bf z}_1,{\bf z}_2)
   \chi^{E}_{\xi^{\prime}\eta}({\bf z}_1,{\bf z}_2) \;,
\label{salpchi}
\end{equation}
where the Salpeter amplitude has been defined by
\begin{equation}
   \chi^{E}_{\alpha\sigma}({\bf x}_1,{\bf y}_2)
   \equiv
   \langle \Psi_0 | \psi_{\alpha}({\bf x}_1)
   \bar{\psi}_{\sigma}({\bf y}_2) | \Psi_E \rangle  \;.
\label{chidef}
\end{equation}
To interpret $\chi^{E}_{\alpha\sigma}$, we expand the fermion fields
in terms of a free single-particle basis:
\begin{equation}
 \psi_{\alpha}({\bf x}) = \sum_{{\bf k}s}
  \Big(
    {\left[U_{{\bf k}s}({\bf x})\right]}_\alpha
    b_{s}({\bf k}) +
    {\left[V_{{\bf k}s}({\bf x})\right]}_\alpha
    {d}^{\dagger}_{s}({\bf k})
  \Big)\;,
\end{equation}
where $b_{s}({\bf k})$ and $d_{s}({\bf k})$ are second-quantized
operators, and $U_{{\bf k}s}({\bf x})$ and $V_{{\bf k}s}({\bf x})$
are free single-particle Dirac spinors (see Appendix~\ref{singpart}).
In this representation, Salpeter's amplitude becomes
\begin{eqnarray}
   \chi^{E}_{\alpha\sigma}({\bf x}_1,{\bf y}_2)
 = \sum_{{\bf k}_{1}s_{1};{\bf k}_{2}s_{2}}
   \bigg(
     &&
     \Big[U_{{\bf k}_{1}s_{1}}({\bf x}_1)\Big]_\alpha
     \Big[\bar{U}_{{\bf k}_{2}s_{2}}({\bf y}_2)\Big]_\sigma
     \langle \Psi_{0} |
       b_{{s}_{1}}({\bf k}_{1})
       b^{\dagger}_{{s}_{2}}({\bf k}_{2})
     | \Psi_{E} \rangle + \nonumber \\
     &&
     \Big[V_{{\bf k}_{1}s_{1}}({\bf x}_1)\Big]_\alpha
     \Big[\bar{V}_{{\bf k}_{2}s_{2}}({\bf y}_2)\Big]_\sigma
     \langle \Psi_{0} |
       d^{\dagger}_{{s}_{1}}({\bf k}_{1})
       d_{{s}_{2}}({\bf k}_{2})
     | \Psi_{E} \rangle + \nonumber \\
     &&
     \Big[U_{{\bf k}_{1}s_{1}}({\bf x}_1)\Big]_\alpha
     \Big[\bar{V}_{{\bf k}_{2}s_{2}}({\bf y}_2)\Big]_\sigma
     \langle \Psi_{0} |
       b_{{s}_{1}}({\bf k}_{1})
       d_{{s}_{2}}({\bf k}_{2})
     | \Psi_{E} \rangle + \nonumber \\
     &&
     \Big[V_{{\bf k}_{1}s_{1}}({\bf x}_1)\Big]_\alpha
     \Big[\bar{U}_{{\bf k}_{2}s_{2}}({\bf y}_2)\Big]_\sigma
     \langle \Psi_{0} |
       d^{\dagger}_{{s}_{1}}({\bf k}_{1})
       b^{\dagger}_{{s}_{2}}({\bf k}_{2})
     | \Psi_{E} \rangle
   \bigg) \;.
\label{chidef2}
\end{eqnarray}
Since the free single-particle solutions are known, all dynamical
information about the two-body system is contained in the four
probability amplitudes $\Big(\langle\Psi_0|b_{{s}_{1}}({\bf k}_{1})
b^{\dagger}_{{s}_{2}}({\bf k}_{2})|\Psi_E\rangle, \ldots\Big)$.
However, because of the instantaneous approximation assumed
for the two-body kernel, two of these amplitudes vanish
\begin{equation}
     \langle \Psi_{0} |
       b_{{s}_{1}}({\bf k}_{1})
       b^{\dagger}_{{s}_{2}}({\bf k}_{2})
     | \Psi_{E} \rangle =
     \langle \Psi_{0} |
       d^{\dagger}_{{s}_{1}}({\bf k}_{1})
       d_{{s}_{2}}({\bf k}_{2})
     | \Psi_{E} \rangle = 0 \;,
\label{uubar}
\end{equation}
leading, in turn, to the following form for Salpeter's relativistic
amplitude
\begin{eqnarray}
   \chi^{E}_{\alpha\sigma}({\bf x}_1,{\bf y}_2)
 = \sum_{{\bf k}_{1}s_{1};{\bf k}_{2}s_{2}}
   \bigg(
     &&
     \Big[U_{{\bf k}_{1}s_{1}}({\bf x}_1)\Big]_\alpha
     \Big[\bar{V}_{{\bf k}_{2}s_{2}}({\bf y}_2)\Big]_\sigma
      B_{s_{1}s_{2}}({\bf k}_{1},{\bf k}_{2}) + \nonumber \\
     &&
     \Big[V_{{\bf k}_{1}s_{1}}({\bf x}_1)\Big]_\alpha
     \Big[\bar{U}_{{\bf k}_{2}s_{2}}({\bf y}_2)\Big]_\sigma
      D_{s_{1}s_{2}}({\bf k}_{1},{\bf k}_{2})
   \bigg) \;.
\label{chidef3}
\end{eqnarray}
Note that we have introduced the following definitions
\begin{eqnarray}
   B_{s_{1}s_{2}}({\bf k}_{1},{\bf k}_{2}) &\equiv&
     \langle \Psi_{0} |
       b_{{s}_{1}}({\bf k}_{1})
       d_{{s}_{2}}({\bf k}_{2})
     | \Psi_{E} \rangle \;, \\
   D_{s_{1}s_{2}}({\bf k}_{1},{\bf k}_{2}) &\equiv&
     \langle \Psi_{0} |
       d^{\dagger}_{{s}_{1}}({\bf k}_{1})
       b^{\dagger}_{{s}_{2}}({\bf k}_{2})
     | \Psi_{E} \rangle \;.
\end{eqnarray}
Finally, one obtains Salpeter's eigenvalue equation in the center
of momentum frame $({\bf k}_{1}=-{\bf k}_{2}\equiv{\bf k})$ by
projecting out the two dynamical amplitudes from
Eq.~(\ref{salpchi})
\begin{eqnarray}
 \left[+E-2E_{k} \right]B_{s_1 s_2}({\bf k})
   =
  \int\frac{d {\bf k}'}{(2\pi)^3}\sum_{{s}_{1}^\prime ,{s}_{2}^\prime}
 &\biggl[&
            \langle{{\bf k};s_1 , s_2}\vert{V^{++}}
             \vert {{\bf k}^\prime;{s}_{1}^\prime,{s}_{2}^\prime} \rangle
             B_{s_1^\prime s_2^\prime}({\bf k}^\prime) \nonumber \\
  &+&       \langle{{\bf k};s_1 , s_2}\vert{V^{+-}}
             \vert {{\bf k}^{\prime};{s}_{1}^{\prime},{s}_{2}^{\prime}}\rangle
             D_{s_1^\prime s_2^\prime}({\bf k}^\prime)
   \biggl] \;,
\label{btilde2}
\end{eqnarray}
\begin{eqnarray}
 \left[-E - 2E_{k} \right] D_{s_1 s_2}({\bf k})
  =
   \int\frac{d {\bf k}'}{(2\pi)^3}\sum_{{s}_{1}^\prime ,{s}_{2}^\prime}
  &\biggl[&
           \langle{{\bf k};s_1 , s_2}\vert{V^{+-}}
            \vert {{\bf k}^\prime;{s}_{1}^\prime,{s}_{2}^\prime} \rangle
             B_{s_1^\prime s_2^\prime}({\bf k}^\prime) \nonumber \\
    &+&    \langle{{\bf k};s_1 , s_2}\vert{V^{++}}
            \vert {{\bf k}^\prime;{s}_{1}^\prime,{s}_{2}^\prime} \rangle
             D_{s_1^\prime s_2^\prime}({\bf k}^\prime)
   \biggl] \;,
\label{dtilde2}
\end{eqnarray}
where the matrix elements of the potential have been defined as
\begin{eqnarray}
 \langle{{\bf k};s_1,s_2}\vert{V^{++}}
  \vert {{\bf k}^\prime;{s}_{1}^\prime,{s}_{2}^\prime} \rangle
 &=& {\left[\bar{U}_{s_1}({\bf k})\right]}
     {\left[\bar{U}_{s_2}(-{\bf k})\right]}
     {V_{\rm C}({\bf k}-{\bf k}')}
     {\left[U_{{s}_{1}^\prime}({\bf k^\prime})\right]}
     {\left[U_{{s}_{2}^\prime}(-{\bf k^\prime})\right]} \;,  \\
\label{vpp}
 \langle{{\bf k};s_1,s_2}\vert{V^{+-}}
  \vert {{\bf k}^\prime;{s}_{1}^\prime,{s}_{2}^\prime} \rangle
 &=& {\left[\bar{U}_{s_1}({\bf k})\right]}
     {\left[\bar{U}_{s_2}(-{\bf k})\right]}
     {V_{\rm C}({\bf k}+{\bf k}')}
     {\left[V_{{s}_{1}^\prime}({\bf k^\prime})\right]}
     {\left[V_{{s}_{2}^\prime}(-{\bf k^\prime})\right]} \;.
\label{vpm}
\end{eqnarray}
This eigenvalue equation is identical to the one derived
in Ref.~\cite{piek92} for the two-fermion case, with the
two-body interaction (in the $F\bar{F}$ case) obtained
from charge conjugation
\begin{equation}
     V_{\rm C} \equiv \cases{
            + V \;, & for scalar, pseudoscalar, axial vector; \cr
            - V \;, & for vector , tensor. \cr}
\end{equation}

The formal derivation of Salpeter's equation was carried out to the
end of identifying its algebraic structure. Note that Salpeter's
equation can be cast in the following compact matrix form:
\begin{eqnarray}
  \left( \begin{array}{rr}
         H^{++}  &  H^{+-}  \\
         -H^{+-} &  -H^{++} \\
         \end{array}
  \right)
  \left( \begin{array}{c}
         {{B}} \\
         {{D}} \\
         \end{array}
  \right) =
E \left( \begin{array}{c}
         {{B}} \\
         {{D}} \\
         \end{array}
  \right) \;,
\label{RPA}
\end{eqnarray}
where the matrix elements of the ``Hamiltonian'' are given by
\begin{eqnarray}
  \langle {\bf k}; s_1, s_2 | H^{++} |
  {\bf k}'; s'_{1},s'_{2} \rangle   &=&
  \langle {\bf k}; s_1, s_2 | V^{++} |
  {\bf k}'; s'_{1},s'_{2} \rangle +
  2E_{k} (2\pi)^3 \delta({\bf k}-{\bf k}^\prime)
  \delta_{s_{1} s_{1}^{'}}\delta_{s_{2} s_{2}^{'}} \;, \\
  \langle {\bf k}; s_1, s_2 | H^{+-} |
  {\bf k}'; s'_{1},s'_{2} \rangle   &=&
  \langle {\bf k}; s_1, s_2 | V^{+-} |
  {\bf k}'; s'_{1},s'_{2} \rangle    \;.
\end{eqnarray}
The central result from the present section is the recognition that
Salpeter's eigenvalue equation, as given by Eq.~(\ref{RPA}), has the
same algebraic structure as an RPA equation~\cite{piek92,long84,resag93}.
The equivalence between the two equations will enable us to use the
stability analysis developed by Thouless, in the context of nuclear
collective excitations~\cite{thou60}, to study the Lorentz structure of
the confinement potential.

It is also useful to study the dynamical content of the different
contributions to Salpeter's equation. For example, the $V^{++}$
term describes the direct scattering between the constituents. In
the nonrelativistic limit, the iteration of this contribution to
all orders results in Schr\"odinger's equation. In contrast,
the $V^{+-}$ term arises from a genuinely relativistic effect,
namely, the coupling between positive- and negative-energy states.
In particular, this term is responsible for generating the double
Z-graphs in time-ordered perturbation theory. Other contributions,
such as single Z-graphs, do not appear in Salpeter's instantaneous
approximation.

We will be comparing our results in three different limits. As
mentioned previously, Salpeter's approximation is obtained, from
the full relativistic Bethe-Salpeter equation, in the limit of
an instantaneous interaction. A simpler limit, given by the Breit
equation, is obtained by neglecting the double Z-graphs from Salpeter's
equation (this limit is known as the Tamm-Dancoff approximation
in the context of nuclear collective excitations). Still, in this limit
one retains the relativistic corrections to the potential and the
relativistic dispersion relation. Moreover, one also incorporates the
full Lorentz structure of the potential. Finally, the nonrelativistic
Schr\"odinger limit is obtained by neglecting all relativistic
corrections. Formally, it can be obtained by enforcing the following
limits:
\begin{eqnarray*}
   i&)& \;
     E_{k} \rightarrow M + \frac{k^2}{2M} \;, \\
  ii&)& \;
     \langle {\bf k};s_1,s_2 | V^{++}
     | {\bf k}';s_{1}^{'},s_{2}^{'} \rangle
     \rightarrow V({\bf k} - {\bf k}')           \;, \\
 iii&)& \;
     \langle {\bf k};s_1,s_2 | V^{+-}
     | {\bf k}';s_{1}^{'},s_{2}^{'} \rangle =
     \langle {\bf k};s_1,s_2 | V^{-+}
     | {\bf k}';s_{1}^{'},s_{2}^{'} \rangle \equiv 0 \;.
\end{eqnarray*}
\subsection{Stability analysis}
\label{secrpa}
Having identified the algebraic (RPA) structure of Salpeter's
equation, we now employ the same formalism developed by Thouless
in his study of nuclear collective excitations~\cite{thou60}.
We perform the stability analysis by using confining potentials
having both scalar and timelike-vector Lorentz structures
\begin{equation}
  V(r) = \sigma r
    \cases{ {\bf 1}_{1}{\bf 1}_{2} \,,       & for scalar;   \cr
            \gamma^{0}_{1}\gamma^{0}_{2}\,,  & for timelike, \cr}
\end{equation}
where $\sigma$ is the string tension, and we note that both potentials
become identical in the nonrelativistic limit. Salpeter's, and in general
any RPA-like, equation can be rewritten as a Hermitian eigenvalue equation
for the square of the energy~\cite{chi70,ullah71}. This implies that
while the square of the energy is guaranteed to be real, the energy
itself might not. The appearance of solutions having $E^{2} < 0$ signals,
in the context of nuclear collective excitations, an instability of the
ground state against the formation of particle-hole pairs --- a collective
mode with imaginary energy can build up indefinitely.  Thouless has shown
that the stability of the nuclear ground state depends on the Hermitian
matrix
\begin{equation}
           \left( \begin{array}{cc}
                    H^{++} &  H^{+-}  \\
                    H^{+-} &  H^{++}  \\
                   \end{array}
            \right) \;,
\end{equation}
being positive-definite, i.e., all its eigenvalues must be
positive~\cite{thou60,chi70,ullah71}. This condition is
equivalent to requiring that, both, the sum and difference
matrices
\begin{eqnarray}
  H^+ &\equiv& \Big( H^{++} + H^{+-} \Big) \;,  \\
  H^- &\equiv& \Big( H^{++} - H^{+-} \Big) \;,
\end{eqnarray}
be positive definite~\cite{chi70}. In this form, the stability
condition of Salpeter's equation is reduced to finding the
eigenvalues of the two Hermitian matrices $H^+$ and $H^-$. The
existence of a single negative eigenvalue, of either $H^{+}$ or
$H^{-}$, is, thus, sufficient to signal the instability. It is
this criterion that we adopt here to test the stability of
Salpeter's equation.

We concentrate our analysis on the pseudoscalar ($J^{\pi}=0^{-}$)
channel. For this case, Salpeter's equation, for the reduced
amplitudes $b(k)\equiv kB(k)$ and $d(k)\equiv kD(k)$, takes
the following form
\begin{eqnarray}
  \left(+E - 2E_{k} \right) b(k)
   &=&
  \int_{0}^{\infty} \frac{d{k^\prime}}{(2\pi)^3}
    \biggl \{
       \langle k | {V}^{++} | k' \rangle b(k') +
       \langle k | {V}^{+-} | k' \rangle d(k')
     \biggl \} \;,
\label{BKPION} \\
  \left(-E - 2E_{k} \right) d(k)
   &=&
    \int_{0}^{\infty} \frac{d{k^\prime}}{(2\pi)^3}
     \biggl \{
       \langle k | {V}^{+-} | k' \rangle b(k') +
       \langle k | {V}^{++} | k' \rangle d(k')
     \biggl \} \;.
\label{DKPION}
\end{eqnarray}
In spite of the simplicity of the angular momentum content of this
channel, the matrix elements of the potential are complicated by
relativistic corrections
\begin{eqnarray}
 \langle k | V^{++} | k' \rangle &=&
 \left( {E_{k}  + M \over 2 E_{k}  }  \right)
 \left( {E_{k'} + M \over 2 E_{k'} }  \right)
  \Bigg\{
   \left[ 1 + \zeta^{2}_{k}\zeta^{2}_{k'} \right] V_0(k,k') \mp
          2\zeta_{k}\zeta_{k'} V_1(k,k')
  \Bigg\}  \;,
  \label{vpionpp} \\
 \langle k | V^{+-} | k' \rangle &=&
 \left( {E_{k}  + M \over 2 E_{k}  }  \right)
 \left( {E_{k'} + M \over 2 E_{k'} }  \right)
  \Bigg\{
   \left[ \zeta^{2}_{k} + \zeta^{2}_{k'} \right] V_0(k,k') \pm
         2\zeta_{k}\zeta_{k'} V_1(k,k')
  \Bigg\}  \;,
  \label{vpionpm}
\end{eqnarray}
where the upper(lower) sign in the above expressions should be
used for scalar(timelike) confinement. We have also introduced
the kinematical variable
\begin{equation}
 \zeta_{k} \equiv {k \over E_{k} + M} \;,
 \label{zeta}
\end{equation}
to quantify the importance of relativity since, in the nonrelativistic
limit ($\zeta_{k},\zeta_{k'} \rightarrow 0$), these expressions reduce
to
\begin{equation}
 \langle k | V^{++} | k' \rangle \rightarrow V_0(k,k') \;, \quad
 \langle k | V^{+-} | k' \rangle \rightarrow 0         \;.
\end{equation}
Thus, the appearance of the ${\cal L}=1$ component of the potential is
a consequence of relativity. The angular-momentum components of the
potential have been defined by
\begin{equation}
  V_{\cal L}(k,k^\prime)
  = (4\pi)^{2}\int_{0}^{\infty} {dr} \;
    \hat{\jmath}_{\cal L}(kr) V(r) \hat{\jmath}_{\cal L}(k^\prime r) \;,
\label{vlkkp}
\end{equation}
with $\hat{\jmath}_{\cal L}(x) \equiv x j_{\cal L}(x)$ being the
Ricatti-Bessel function.  For confining potentials, the above
integral is ill-defined. Hence, in examining confinement
in momentum space we employ the following regularization for the
spatial part of the potential~\cite{spence86,maung93}:
\begin{equation}
  V(r) = \sigma r e^{-\eta r} \equiv
         \sigma \frac{\partial^{2}}{\partial\eta^{2}}
         \frac{e^{-\eta r}}{r}  \;.
\end{equation}
The Fourier transform of the potential is now well behaved
and is given by
\begin{equation}
    V({\bf k}-{\bf k}') =
    \frac{\partial^{2}}{\partial\eta^{2}}
    \left[
      \frac{4\pi\sigma}{({\bf k}-{\bf k}')^2 + \eta^2}
    \right] \;.
\label{vkansatz}
\end{equation}
Evidently, we are interested in studying the stability of Salpeter's
equation in the limit of $\eta \rightarrow 0$.
The stability analysis requires the explicit evaluation of
$V^{+}$ and $V^{-}$. These are computed with the help of
Eqs.~(\ref{vpionpp}) and~(\ref{vpionpm})
\begin{eqnarray}
  V^{+}(k,k') &\equiv&
  \langle k | V^{++} + V^{+-} | k' \rangle = V_0(k,k') \;,
 \label{vplus} \\
  V^{-}(k,k') &\equiv&
  \langle k | V^{++} - V^{+-} | k' \rangle = V_0(k,k')\xi(k,k') \;,
 \label{vminus}
\end{eqnarray}
where we have introduced relativistic ``correction'' factors,
separately, for scalar and timelike confinement
\begin{eqnarray}
 \xi_{\rm s}(k,k') &\equiv&
  \left[
   {M^{2} \over E_{k}E_{k'}} -
   {kk'   \over E_{k}E_{k'}}
   {V_1(k,k') \over V_0(k,k')}
  \right] \;,
 \label{xis} \\
 \xi_{\rm v}(k,k') &\equiv&
  \left[
   {M^{2} \over E_{k}E_{k'}} +
   {kk'   \over E_{k}E_{k'}}
   {V_1(k,k') \over V_0(k,k')}
  \right] \;.
 \label{xiv}
\end{eqnarray}
Note that these correction factors are different because of an
important relative minus sign. The expression for $V^{+}$ is
remarkably simple --- no vestige of relativistic effects remain.
The only relativistic corrections to $H^{+}$ are, thus, kinematical
and fully contained in the kinetic-energy operator $E_{k}$. The
eigenvalue equation for $H^{+}$ is, then, a simple
``nonrelativistic'' Schr\"odinger equation with a relativistic
kinetic-energy term:
\begin{equation}
  \left(E_{NR} - 2E_{k} \right) b_{NR}(k) =
  \int_{0}^{\infty} \frac{d{k^\prime}}{(2\pi)^3}
   V_{0}(k,k') b_{NR}(k') \;.
  \label{hplus}
\end{equation}
Since the lowest eigenvalue of $H^{+}$ is at least as large as
twice the constituent mass, we conclude that, if present, the
instability must arise from the relativistic effects on $H^{-}$.

The relativistic corrections to $V^{-}$ are contained in the
expressions given in Eqs.~(\ref{xis}) and~(\ref{xiv}). In
addition to simple kinematical factors, these terms contain
important dynamical corrections which depend on the particular
details of the potential. In general, these expressions are
complicated. Yet, there is one limit for which they are simple
and illuminating. This is the $k'=k$ limit for which the singularity
structure of the potential --- and, thus, the dynamical realization
of confinement --- become manifest. Separately, $V_{0}$ and $V_{1}$ are
singular in the $\eta \rightarrow 0$ limit:
\begin{eqnarray}
  V_{0}(k,k'=k) &=& {8\pi^{2}\sigma \over \eta^{2}}
                    [1 + {\cal O}(\eta^2/k^2)]  \;,
  \label{vzero}                                 \\
  V_{1}(k,k'=k) &=& {8\pi^{2}\sigma \over \eta^{2}}
                    \left[1 -
                     \displaystyle{
                      {\eta^{2} \over 2k^{2}}
                       \ln\left({\eta^{2} \over 4k^{2}}\right)} +
                    {\cal O}(\eta^2/k^2)\right] \;.
  \label{vone}
 \label{limits}
\end{eqnarray}
Yet, the leading singularities cancel in forming the ratio
\begin{equation}
  \lim_{\eta\rightarrow 0}
  {V_{1}(k,k'=k) \over V_{0}(k,k'=k)} =
  \lim_{\eta\rightarrow 0}
  \left[1 - \displaystyle{
        {\eta^{2} \over 2k^{2}}
        \ln\left({\eta^{2} \over 4k^{2}}\right)} +
        {\cal O}(\eta^2/k^2)\right] = 1 \;.
 \label{limit}
\end{equation}
This result generates the following simple corrections to the
nonrelativistic potential
\begin{eqnarray}
 \xi_{\rm s}(k,k) &=&
   {M^{2} - k^{2} \over M^{2} + k^{2}} \rightarrow
   \cases{+1 \;, & if $k/M \rightarrow 0     $;  \cr
          -1 \;, & if $k/M \rightarrow \infty$,  \cr}   \\
 \xi_{\rm v}(k,k) &=&
   {M^{2} + k^{2} \over M^{2} + k^{2}} = 1 \;.
\end{eqnarray}
Note that in spite of their simplicity, these expressions
dictate how the singularity structure of the potential
is modified by relativistic effects.

For timelike confinement, the singularity structure of $V^{-}$
remains unchanged from its nonrelativistic value. In particular,
for $k'=k$ we obtain [as in Eq.~(\ref{vzero})]
\begin{equation}
  V^{-}_{\rm v}(k,k) \equiv V_{0}(k,k)\xi_{\rm v}(k,k)
                     = {8\pi^{2}\sigma \over \eta^{2}}
                       [1 + {\cal O}(\eta^2/k^2)] \;.
\end{equation}
Away from the ($k'=k$) singularity the potential receives
corrections from relativity. While these corrections are
quantitatively significant, they do not lead to any important
changes in the qualitative behavior of the potential. These
small changes have been documented in Table~\ref{table1}, and
also in Fig.~\ref{fig1}, and support the assertion that a
confining potential having a timelike Lorentz structure does not
generate an instability. The picture is, however, drastically
different for scalar confinement (see Fig.~\ref{fig2}). In this
case the singularity structure of the potential becomes
\begin{equation}
  V^{-}_{\rm s}(k,k) \equiv V_{0}(k,k)\xi_{\rm s}(k,k)
                     = {8\pi^{2}\bar{\sigma}(k^{2}) \over \eta^{2}}
                      [1 + {\cal O}(\eta^2/k^2)] \;,
\end{equation}
where we have introduced a momentum-dependent string tension
\begin{equation}
   \bar{\sigma}(k^{2}) \equiv \sigma
   \left({1 - k^{2}/M^{2} \over 1 + k^{2}/M^{2}}\right) \rightarrow
   \cases{+\sigma \;, & if $k/M \rightarrow 0     $;  \cr
          -\sigma \;, & if $k/M \rightarrow \infty$.  \cr}
\end{equation}
The presence of this momentum-dependent string tension implies
that, for momenta larger than the constituent quark mass, the
potential shifts from a ``rising'' ($\bar{\sigma} > 0$) into a
``sliding'' ($\bar{\sigma} < 0$) regime. Moreover, this behavior
is unavoidable --- provided one can mix in large-enough momentum
components. However, if one limits the analysis to a small number
of basis states, thus effectively introducing a momentum cutoff,
the sliding regime will be missed --- and so will the instability
--- whenever the constituent mass exceeds the value of the cutoff.
This fact is clearly displayed in Table~\ref{table2}. For a quark
mass of $M=0.9$~GeV, it is only after including 25 basis states that
one can uncover the instability.

We conclude this section by considering a mixture of scalar
and timelike confinement, i.e., the Lorentz structure of the
potential is assumed to be
\begin{equation}
  \Gamma \equiv x\gamma^{0}_{1}\gamma^{0}_{2} +
                (1-x) {\bf 1}_{1}{\bf 1}_{2}  \;,
\end{equation}
where $x$ denotes the fraction of timelike structure. In this case
\begin{eqnarray}
  V^{+}(k,k') &=& V_0(k,k') \;,  \\
  V^{-}(k,k') &=&
  \left[
         {M^{2} \over E_{k}E_{k'}} +
   (2x-1){kk'   \over E_{k}E_{k'}}
   {V_1(k,k') \over V_0(k,k')}
  \right] \;.
\end{eqnarray}
As before, $V^{+}$ is insensitive to relativistic correction
and $H^{+}$ remains positive definite. In contrast, $H^{-}$ is
positive definite only for $x\ge 1/2$. Hence, any mix of
scalar and timelike Lorentz structures has stable solutions
only for $x$ in the interval $0.5\le x\le 1$. This result is
in agreement with the numerical evidence presented in
Ref.~\cite{arch93}.

\section{Results}
\label{secresults}
For one-boson exchange potentials, like those encountered in
the nucleon-nucleon problem, one can evaluate Eqs.~(\ref{BKPION})
and~(\ref{DKPION}) quite efficiently in momentum space using a
Gauss quadrature scheme~\cite{piek92,maung93}. However, it has
been recently suggested that this representation is not appropriate
for the case of a confining potential~\cite{maung93} due to its
highly singular structure in momentum space. Evidently, the confining
potential is most easily treated in configuration space. In contrast,
the relativistic kinetic energy operator is difficult to handle in
configuration space but easily treated in momentum space. We can
accommodate both of these requirements by expanding Salpeter's equations
in a suitable basis. Here, we use the radial eigenfunctions of the
nonrelativistic harmonic oscillator, $R_{nL}$, to expand the two
amplitudes $B_{LSJ}$ and $D_{LSJ}$ in terms of unknown coefficients
\begin{eqnarray}
  B_{LSJ}(k) &=& \sum_{n} B_{nLSJ} R_{nL}(k) \\
  D_{LSJ}(k) &=& \sum_{n} D_{nLSJ} R_{nL}(k) \;.
\end{eqnarray}
This procedure results in a matrix equation for the unknown
coefficients $B_{nLSJ}$ and $D_{nLSJ}$ which can be diagonalized
using the method developed by Ullah and Rowe~\cite{ullah71}. Upon
diagonalization one obtains the (previously) unknown coefficients
from which one then can reconstruct the two amplitudes $B_{LSJ}$ and
$D_{LSJ}$, and, ultimately, the Salpeter wave function
${\raise 2pt \hbox{$\chi$}}_{J^\pi}$.

We have made an effort to customize our numerical code so that it can
be readily compared with previous results obtained from various
approximations to the Bethe-Salpeter equation~\cite{jacobs86,fulcher93}.
Specifically, mass spectra obtained from the Salpeter code have been
checked against nonrelativistic (Schr\"odinger) results and in the
so-called ``spinless-Salpeter'' limit where a relativistic dispersion
relation has been employed. Comparisons have also been made to the
results obtained by Long~\cite{long84}, and Spence and
Vary~\cite{spence86}, in the Breit limit where the coupling to the
negative-energy states is ignored ($V^{+-} \equiv 0$). These
comparisons were effected using a static potential consisting
of a OGE piece plus a scalar confinement. We note that since
these approximations neglect the coupling to negative energy states,
they are not suitable to uncover the instability. Our results are
also in agreement with a recent numerical study of the meson spectrum
using the full Bethe-Salpeter equation (with a static interaction)
in which it was found that the smallest possible eigenvalue decreases
until it becomes imaginary as one increases the dimension of the
basis~\cite{munz93}.

We now proceed to show results for the meson spectra using a
variety of constituent quark masses. Since the main goal of
this work is the study of the Lorentz structure of the confinement
potential, rather than a fit to the experimental spectrum, we ignore
the one-gluon-exchange component of the potential. In Fig.~\ref{fig3}
we show the square of the ground-state energy in the pseudoscalar
($J^{\pi}=0^{-}$) channel for a string-tension of
$\sigma=0.29$~GeV$^{2}$ and a constituent mass of $M=300$~MeV.
Results are presented as a function of the oscillator parameter
$\beta$ for various sizes of the basis. We have indicated
with ``pluses'' the results obtained by assuming a confinement
potential having a timelike Lorentz structure. In this case,
the development of a well-defined plateau signals the stability
of the solution. In contrast, the presence of an instability in
the scalar case (shown with ``diamonds'') is manifested by the
appearance of imaginary eigenvalues ($E^{2}<0$) --- even after
including only five basis states.

The presence of imaginary eigenvalues is only one of many
possible manifestations of an instability. In particular,
at the onset of the instability (i.e., for $E = 0$) the solution
is also characterized by having $B_{LSJ} =\pm D_{LSJ}$ and, thus,
zero norm. In Fig.~\ref{fig4} we have plotted the two
(reduced) pseudoscalar amplitudes $b$ and $d$ as a function of the
relative separation between the quarks using $\beta=0.85$~GeV.
This represents a value of the oscillator parameter for which
the energy is about to become imaginary. The approximate validity
of the $b=-d$ relation is evident. In addition, the behavior of
both amplitudes, as well as the relative importance of the different
basis states, suggests that the solution is, indeed, mixing in basis
states having large momentum components. This result is in striking
contrast with the one presented in Fig.~\ref{fig5} for timelike
confinement. Not only do both amplitudes display a very smooth behavior
with $r$, but, from the relative size of the amplitudes we can conclude
that the relativistic corrections are small down to values of the
quark mass of $M=300$~MeV.

In Figs.~\ref{fig6},\ref{fig7}, and \ref{fig8} we have carried
out the same analysis as above but for a larger value of the
constituent mass, i.e., $M=900$~MeV. The main purpose of
Fig.~\ref{fig6} is to indicate how easily one can be mislead into
believing that the solution for scalar confinement is stable. Limiting
the calculation to 5 or 15 basis states results in a smooth and stable
behavior for the energy. Yet, as soon as basis states having
large-enough momentum components are mixed in, the instability becomes
manifest. The striking difference displayed in Fig.~\ref{fig7} for the
behavior of the amplitudes as a function of the number of basis states,
serves to further reinforce our conclusion. Finally, Fig.~\ref{fig8}
shows the stability of the timelike solution against an increase in the
number of basis states.
\section{Conclusions}
\label{secconcl}
We have investigated the Lorentz structure of the confinement
potential by means of a stability analysis of the meson spectra
using Salpeter's instantaneous approximation to the Bethe-Salpeter
equation. Because the algebraic structure of Salpeter's equation
is identical in form to an RPA equation, we benefited from the
techniques developed by Thouless in his study of nuclear collective
excitations. The analysis revealed an instability, manifested by the
appearance of imaginary eigenvalues, for a confinement potential
assumed to transform as a Lorentz scalar. Moreover, an analytic study
suggests that this instability will persist even for large values of
the constituent mass --- provided one can mix in large-enough momentum
components. We have explicitly demonstrated that if, instead, the
calculation is limited to a small number of basis states, one can
conclude --- erroneously --- that the stability will get restored for
large-enough values of the quark mass. In essence, we have shown that
for momentum components larger than the constituent mass, relativistic
effects modify the singularity structure of the potential and give rise,
effectively, to a negative string tension.

In contrast to the scalar case, we have found no instability for
timelike confinement. Furthermore, a study of the lowest mass
pseudoscalar state suggests small relativistic corrections down to
values of the quark mass of $M=300$~MeV.

Many issues remain to be addressed. For example, how sensitive
are our results to the choice of an instantaneous approximation.
The answer to this question is difficult because very little,
if at all, is known about the dynamical origin of the confinement
potential. Thus, throughout this work we have adopted the same static
form for the confining potential revealed to us from lattice gauge
studies. Still, the role of non-instantaneous contributions, such
as retardation, remains an important open problem.

A natural extension of this work could consist of a detailed study
of the meson spectra using a two-body interaction having a timelike
confining piece plus a one-gluon-exchange contribution. This Lorentz
choice for the confinement is dictated, not only by the previous
stability analysis, but, also, by the correct reproduction of the
linear Regge trajectories~\cite{lagae92,munz93}. Unfortunately,
a timelike confinement potential leads to a relativistic spin-orbit
contribution of the ``wrong'' sign~\cite{lagae92}. Indeed, the sign
of the spin-orbit splitting seems to constitute our best evidence
in support of scalar confinement~\cite{gromes91}. One could attempt
to remedy this deficiency by introducing a mixture of scalar plus
timelike confinement~\cite{arch93} or some other imaginative
solution~\cite{lagae92}. Regardless of the final outcome, it should
be evident that the long-range part of the $q\bar{q}$ interaction
will remain at the center of many future investigations.

\acknowledgments

We thank C.~Long, C.~M\"unz, D.~Robson, and A.~Williams for many
helpful discussions. This research was supported by the Florida
State University Supercomputer Computations Research Institute
and the U.S. Department of Energy contracts DE-FC05-85ER250000,
DE-FG05-92ER40750, and DE-FG05-86ER40273.
\appendix
\section{Single-particle expansion of fermion fields}
\label{singpart}
The concept of a particle-antiparticle removal amplitude is basis-dependent;
one is asking what is the probability of removing a pair having, for example,
momenta ${\bf k}_1$ and ${\bf k}_2$, from the system.  To address this, the
fermion-field operators can be expanded in terms of an (initially) unspecified
basis:
\begin{equation}
 \psi_{\alpha}({\bf x}) = \sum_{i}
 \biggl \{ {\left[U_{i}({\bf x})\right]}_\alpha b_{i}
     +    {\left[V_{i}({\bf x})\right]}_\alpha {d}^{\dagger}_{i}
 \biggl \} .
\end{equation}
The second-quantized operators $b_i , d_{i}^{\dagger}$ respectively annihilate
a particle or create an antiparticle in the corresponding single-particle
state; they satisfy the usual canonical anticommutation relations.
$U,V$ are solutions to a single-particle Dirac Hamiltonian; they are
orthonormal
\begin{eqnarray}
   \int {d^3\! x} U_{i}^{\dagger}({\bf x}) U_{j}({\bf x})
 &=& \int {d^3\! x} V_{i}^{\dagger}({\bf x}) V_{j}({\bf x})
  = \delta_{ij} , \nonumber \\
   \int {d^3\! x} U_{i}^{\dagger}({\bf x}) V_{j}({\bf x})
 &=& \int {d^3\! x} V_{i}^{\dagger}({\bf x}) U_{j}({\bf x})
  = 0 ,
\end{eqnarray}
and satisfy the completeness relation
\begin{equation}
  \sum_{i} \left [ U_{i}({\bf x}) U_{i}^{\dagger}({\bf y})
             +     V_{i}({\bf x}) V_{i}^{\dagger}({\bf y})
           \right ] = \delta({\bf x}-{\bf y}) {\bf 1} .
\end{equation}
We make use of the eigenstates of the free Dirac Hamiltonian, here defined as
\begin{equation}
  U_{{\bf k}s}({\bf x}) = e^{i{\bf k}\cdot{\bf x}}U({\bf k},s)  \;; \quad
  V_{{\bf k}s}({\bf x}) = e^{-i{\bf k}\cdot{\bf x}}V({\bf k},s) \;,
\end{equation}
where
\begin{eqnarray}
 U({\bf k},s)=
 {\left[ \frac{E_{k} + M}{2E_{k}} \right]}^\frac{1}{2}
                \left ( \begin{array}{c}
                        1  \\
                        \displaystyle{
                         \frac{{\bf \sigma}\cdot{\bf k}}{E_{k} + M} }
                        \end{array}
                \right )
                \chi_s \;,   \nonumber \\
                                    \\
\label{freedirac}
 V({\bf k},s)=
 {\left[ \frac{E_{k} + M}{2E_{k}} \right]}^\frac{1}{2}
                \left ( \begin{array}{c}
                        \displaystyle{
                     \frac{{\bf \sigma}\cdot{\bf k}}{E_{k} + M} }  \\
                              1
                        \end{array}
                \right )
                \tilde{\chi}_s \;,   \nonumber
\end{eqnarray}
are positive and negative energy plane-wave spinors,
$\tilde{\chi}_s \equiv {(-)}^{\frac{1}{2}+s}\chi_{-s}$,
with $\chi_s$ the conventional two-component Pauli spinors
\begin{equation}
 \chi_{\frac{1}{2}} = \left( \begin{array}{c}
                                  1 \\
                                  0
                             \end{array}
                      \right)  \;; \quad
 \chi_{-\frac{1}{2}} = \left( \begin{array}{c}

                                   0 \\
                                   1
                              \end{array}
                       \right)  \;.
\end{equation}
\section{Partial-wave decomposition}
\label{angmombasis}
With interactions that conserve total angular momentum and parity, it
becomes convenient to rewrite Eqs.(\ref{btilde2}) and (\ref{dtilde2})
in an angular momentum basis in which the Pauli spins $s_1 , s_2$ of the
two particles are coupled to form a total $S$, then coupling the relative
angular momentum of the two particles $L$ to $S$ to form the total angular
momentum $J$ of the bound state. One can then solve the equation separately
for each different total angular momentum $J$ and parity
$\pi = {(-1)}^{L+1}$. We will illustrate the partial-wave decomposition
of the two-body amplitude with the direct term $V^{++}$ [Eq.~(\ref{vpp})].
Including Dirac indices, this term
is
\begin{equation}
 \langle{{\bf k};s_1,s_2}
 \vert{V^{++}} \vert {{\bf k}^\prime;{s}_{1}^\prime,{s}_{2}^\prime}\rangle
 = {\left[\bar{U}_{s_1}({\bf k})\right]}_\alpha
     {\left[\bar{U}_{s_2}(-{\bf k})\right]}_\beta
     {\left[
     V(\mid{\bf k}-{\bf k}^\prime \mid)
      \right]}_{\alpha\beta ; \alpha^\prime \beta^\prime}
     {\left[U_{{s}_{1}^\prime}({\bf k^\prime})\right]}_{\alpha^\prime}
     {\left[U_{{s}_{2}^\prime}(-{\bf k^\prime})\right]}_{\beta^\prime} \;,
\end{equation}
where the Fourier transform of the potential is given by
\begin{equation}
 V({\bf k}-{\bf k}^\prime)
 = \int d{\bf r} \;
   e^{-i\left({\bf k}-{\bf k}^\prime\right)\cdot{\bf r}}V({\bf r})
   \equiv \sum_{{\cal{L}}M_{\cal{L}}}
   Y_{{\cal{L}}M_{\cal{L}}}(\hat{k})
   \left [ \frac{1}{kk^\prime}{V}_{\cal{L}}(k,k^\prime) \right ]
   Y_{{\cal{L}}M_{\cal{L}}}^{*}(\hat{k}^\prime) \;.
\end{equation}
In order to construct states of good total angular momentum, we write the
free two-body state as a direct product of Pauli spinors, i.e.,
\begin{eqnarray}
  {\left[U_{{s}_{1}}({\bf k})\right]}_{\alpha}
  {\left[U_{{s}_{2}}(-{\bf k})\right]}_{\beta}
 &=&
  {\left[ \frac{E_{k} + M}{2E_{k}} \right]}
   {\left ( \begin{array}{c}
                1  \\
                \displaystyle{
                \frac{{\bf \sigma}\cdot{\bf k}}{E_{k} + M} }
           \end{array}
    \right )}_{\alpha}
   {\left ( \begin{array}{c}
                1  \\
                \displaystyle{
                \frac{-{\bf \sigma}\cdot{\bf k}}{E_{k} + M} }
           \end{array}
    \right )}_{\beta}
    \vert {s_1 s_2} \rangle \nonumber \\
 &=&
   C_{\alpha\beta}(k)
   \sum_{\lambda} \langle{\alpha 0 ; \beta 0}\vert{\lambda 0}\rangle
   {\left[ Y_{\lambda}(\hat{k})
    {\left(\sigma_\alpha \sigma_\beta \right)}_{\lambda}
    \right]}_{0,0}  \vert {s_1 s_2} \rangle ,
\label{directprod}
\end{eqnarray}
where $C_{\alpha\beta}(k)$ is defined as
\begin{equation}
  C_{\alpha\beta}(k)
  = \sqrt{4\pi} {(-1)}^\alpha
    {\left[ \frac{E_{k} + M}{2E_{k}} \right]}
    \xi_{\alpha}(k) \xi_{\beta}(k) \; ; \;
    \xi_{\alpha}(k) = \left \{ \begin{array}{ll}
                          1 & \mbox{if $\alpha$ = 0 ;} \\
                     \frac{k}{E_{k} + M} & \mbox{if $\alpha$ = 1.}
                               \end{array}
                      \right.
\end{equation}
Equation (\ref{directprod}) can now be combined with the partial-wave
expansion of the Salpeter amplitudes [Eqs.~(\ref{bwave})] to give
\begin{equation}
  \sum_{s_1 s_2}
  {\left[U_{{s}_{1}}({\bf k})\right]}_{\alpha}
  {\left[U_{{s}_{2}}(-{\bf k})\right]}_{\beta}
   B_{s_1 s_2}({\bf k})
 = \sum_{LS{\cal L}{\cal S}JM}
   {\cal F}_{{\cal L}{\cal S};L S J}^{\alpha \beta}(k)
   B_{LSJ} \langle{\hat{k}}\vert{{\cal L}{\cal S}JM}\rangle ,
\end{equation}
where we have defined
\begin{equation}
   {\cal F}_{{\cal L}{\cal S};L S J}^{\alpha \beta}(k)
   = C_{\alpha\beta}(k)
     \sum_{\lambda}
     \langle{\alpha 0 ; \beta 0}\vert{\lambda 0}\rangle
     \langle{{\cal L}{\cal S}J}\vert\vert
     {\left
     [Y_{\lambda}
     {\left(\sigma_\alpha \sigma_\beta \right)}_{\lambda}\right]}_0
     \vert\vert{LSJ}\rangle.
\end{equation}
Salpeter's equations in momentum space are then written in the
angular momentum basis as
\begin{eqnarray}
  \left[ +E - 2E_{k} \right] b_{LSJ}(k)
   =
  \int_{0}^{\infty} \frac{d{k^\prime}}{(2\pi)^3}
  \sum_{L^\prime S^\prime}
   & \biggl \{ &
       \langle{k;LSJ}
       \vert {V}^{++} \vert {k^\prime;L^\prime S^\prime J} \rangle
                   b_{L^\prime S^\prime J}(k^\prime) \nonumber \\
     &+& \langle{k;LSJ}
       \vert {V}^{+-} \vert {k^\prime;L^\prime S^\prime J} \rangle
                   d_{L^\prime S^\prime J}(k^\prime)
     \biggl \} \;,
\label{BKLSJ}
\end{eqnarray}
\begin{eqnarray}
  \left[ -E - 2E_{k} \right] d_{LSJ}(k)
   =
  \int_{0}^{\infty} \frac{d{k^\prime}}{(2\pi)^3}
  \sum_{L^\prime S^\prime}
   & \biggl \{ &
       \langle{k;LSJ}
       \vert {V}^{-+} \vert {k^\prime;L^\prime S^\prime J} \rangle
                   b_{L^\prime S^\prime J}(k^\prime)   \nonumber \\
   &+& \langle{k;LSJ}
       \vert {V}^{--} \vert {k^\prime;L^\prime S^\prime J} \rangle
                   d_{L^\prime S^\prime J}(k^\prime)
     \biggl \} \;,
\label{DKLSJ}
\end{eqnarray}
with $b(k) \equiv kB(k)$ and $d(k) \equiv kD(k)$.  The partial-wave
expansions of the Salpeter amplitudes $B$ and $D$ are given by
\begin{eqnarray}
  B_{LSJ}(k)
  &=& \sum_{ M_L M_S}
      \langle {L M_L ; S M_S} \vert {J M} \rangle
      \int {d\hat{k}} Y_{LM_L}^{*}(\hat{k}) B_{S M_S}({\bf k}) \;, \\
\label{bwave}
  D_{LSJ}(k)
  &=& \sum_{ M_L M_s}
      \langle {L M_L ; S M_S} \vert {J M} \rangle
      \int {d\hat{k}} Y_{LM_L}^{*}(\hat{k}) D_{S M_S}({\bf k}) \;, \\
\label{dwave}
\end{eqnarray}
where we have used the following definitions:
\begin{equation}
  B_{S M_S}({\bf k})
  \equiv
  \sum_{s_1 s_2} \langle {s_1 s_2} \vert {S M_S}
  \rangle B_{s_1 s_2}({\bf k}) \;,
\end{equation}
\begin{equation}
  D_{S M_S}({\bf k})
  \equiv
  \sum_{s_1 s_2}
  \langle {s_1 s_2}\vert{S M_S}
  \rangle{(-)}^{1-s_1 -s_2} D_{-s_1 -s_2}({\bf k}) \;.
\end{equation}
The amplitudes $B_{LSJ}$ and $D_{LSJ}$ represent the probability
amplitude of removing and adding a particle-antiparticle pair to
or from the system, with relative momentum $k$; within the framework
of Salpeter's approach, they contain all dynamical information about
the nature of the two-body bound state. The Salpeter wave function
$\chi$ in the angular-momentum basis is then given by
\begin{equation}
   {\left [ {\chi}_{\alpha\beta}^{F\bar{F}}(k) \right ]}_{{\cal L S}J} =
   \sum_{L S}
   \biggl \{ {\cal{F}}_{L{\cal{S}};{\cal{L}}SJ}^{\alpha\bar{\beta}}(k)
            B_{LSJ}(k)
 + {(-)}^{\cal{L}}{\cal{F}}_{L{\cal{S}};{\cal{L}}SJ}^{\bar{\alpha}\beta}(k)
            D_{LSJ}(k)
   \biggl \}  \;.
\label{chilsj}
\end{equation}

For local interactions, the matrix elements of the potential are given
by (a sum over greek indices is implicitly assumed, and
$\bar{\alpha} \equiv 1-\alpha$)
\begin{eqnarray}
   \langle{k;LSJ}
   \vert {V}^{++} \vert {k^\prime;L^\prime S^\prime J} \rangle
 = \langle{k;LSJ}
   \vert {V}^{--} \vert {k^\prime;L^\prime S^\prime J} \rangle
   \nonumber \\
 = \sum_{{\cal L}{\cal S}}
   {(-1)}^{\alpha + \beta} {\cal F}_{{\cal L}{\cal S};LSJ}^{\alpha\beta}(k)
   \langle {{\cal{S}}} \vert\vert
   {\left[V_{\cal{L}}(k,k^\prime)
     \right]}_{\alpha\beta ;{\alpha^\prime}{\beta^\prime}}
   \vert\vert {{\cal{S}}} \rangle
   {\cal F}_{{\cal L}{\cal S};L^\prime S^\prime J}^
   {\alpha^\prime \beta^\prime}(k^\prime) \;,
\end{eqnarray}
\begin{eqnarray}
   \langle{k;LSJ}
   \vert {V}^{+-} \vert {k^\prime;L^\prime S^\prime J} \rangle
 = \langle{k;LSJ}
   \vert {V}^{-+} \vert {k^\prime;L^\prime S^\prime J} \rangle
   \nonumber \\
 = \sum_{{\cal L}{\cal S}}
   {(-1)}^{\alpha + \beta + {\cal L}}
   {\cal F}_{{\cal L}{\cal S};LSJ}^{\alpha\beta}(k)
   \langle {{\cal{S}}} \vert \vert
   {\left[V_{\cal{L}}(k,k^\prime)
     \right]}_{\alpha\beta ;{\alpha^\prime}{\beta^\prime}}
   \vert\vert {{\cal{S}}} \rangle
   {\cal F}_{{\cal L}{\cal S};L^\prime S^\prime J}^
   {\bar{\alpha}^\prime \bar{\beta}^\prime}
          (k^\prime) \;.
\end{eqnarray}
The quantum numbers $L,S$ range only over the values allowed by
$J^\pi$, while $\cal{L},\cal{S}$ can take on all values allowed
by the coupling to $J$.  For $E > 0$, the amplitudes $b,d$ satisfy
the RPA normalization condition \cite{ullah71}
\begin{equation}
  \sum_{L S} \int_{0}^{\infty} \frac{dk}{(2\pi)^3}
   \left [ {b}_{LSJ}^{2}(k) - {d}_{LSJ}^{2}(k) \right ] = 1 .
\label{rpanorm}
\end{equation}
%
%

%
\begin{figure}
\caption{The relativistic correction factor $\xi_{\rm v}(k,k')$
         in the limit of $\eta \rightarrow 0$ for a quark mass
         of $M$=1.0~GeV.}
\label{fig1}
\end{figure}
\begin{figure}
\caption{The relativistic correction factor $\xi_{\rm s}(k,k')$
         in the limit of $\eta \rightarrow 0$ for a quark mass
         of $M$=1.0~GeV.}
\label{fig2}
\end{figure}
\begin{figure}
\caption{Square of the bound-state mass vs the oscillator
         parameter ($\beta$) for a quark mass of $M$=0.3~GeV
         and a string tension of $\sigma$=0.29~${\mbox{GeV}}^2$.
         Results are presented for 5 (dotted line), 15 (dashed line),
         and 25 (solid line) basis states using scalar (diamonds)
         and timelike (pluses) Lorentz structures.}
\label{fig3}
\end{figure}
\begin{figure}
\caption{Salpeter's $b(k)$ (solid line) and $d(k)$ (dashed line)
         amplitudes for scalar confinement using a quark mass of
         $M$=0.3~GeV and a string tension of $\sigma$=0.29~${\mbox{GeV}}^2$.
         Results are presented for 25 basis states using an oscillator
         parameter of $\beta$=0.85 GeV.}
\label{fig4}
\end{figure}
\begin{figure}
\caption{Salpeter's $b(k)$ (solid line) and $d(k)$ (dashed line)
         amplitudes for timelike confinement using a quark mass of
         $M$=0.3~GeV and string tension of $\sigma$=0.29~${\mbox{GeV}}^2$.
         Results are presented for 25 basis states using an oscillator
         parameter of $\beta$=0.70 GeV.}
\label{fig5}
\end{figure}
\begin{figure}
\caption{Square of the bound-state mass vs the oscillator
         parameter ($\beta$) for a quark mass of $M$=0.9~GeV
         and a string tension of $\sigma$=0.29~${\mbox{GeV}}^2$.
         Results are presented for 5 (dotted line), 15 (dashed line),
         and 25 (solid line) basis states using scalar (diamonds)
         and timelike (pluses) Lorentz structures.}
\label{fig6}
\end{figure}
\begin{figure}
\caption{Salpeter's $b(k)$ (solid line) and $d(k)$ (dashed line)
         amplitudes for scalar confinement using a quark mass of
         $M$=0.9~GeV and string tension of $\sigma$=0.29~${\mbox{GeV}}^2$.
         Results are presented for 15 and 25 basis states using an oscillator
         parameter of $\beta$=0.30 GeV.}
\label{fig7}
\end{figure}
\begin{figure}
\caption{Salpeter's $b(k)$ (solid line) and $d(k)$ (dashed line)
         amplitudes for timelike confinement using a quark mass of
         $M$=0.9~GeV and string tension of $\sigma$=0.29~${\mbox{GeV}}^2$.
         Results are presented for 15 and 25 basis states using an oscillator
         parameter of $\beta$=0.30 GeV.}
\label{fig8}
\end{figure}
%
\mediumtext
 \begin{table}
  \caption{Square of the pseudoscalar bound-state mass $E^2$, and
           the lowest eigenvalue $\lambda^{\pm}$ of the matrix $H^{\pm}$
           as a function of the number of basis states for various
           values of the mass. The calculation were done
           assuming timelike confinement and an oscillator parameter
           of $\beta$=0.25 GeV.}
\begin{tabular}{ccccc}
   Mass [MeV]  & $\#$ states & $E^2$ $[\mbox{GeV}^2]$ & $\lambda^+$[GeV]
               &  $\lambda^-$ [GeV]        \\ \hline
    300  & 5  & 4.409  & 1.925  & 2.287    \\
         & 10 & 4.362  & 1.912  & 2.278    \\
         & 15 & 4.358  & 1.911  & 2.277    \\
         & 20 & 4.357  & 1.911  & 2.277    \\
         & 25 & 4.357  & 1.911  & 2.277    \\ \hline
    500  & 5  & 5.330  & 2.195  & 2.428    \\
         & 10 & 5.262  & 2.177  & 2.417    \\
         & 15 & 5.256  & 2.175  & 2.416    \\
         & 20 & 5.255  & 2.175  & 2.416    \\
         & 25 & 5.255  & 2.175  & 2.416    \\ \hline
    900  & 5  & 8.413  & 2.844  & 2.958    \\
         & 10 & 8.265  & 2.814  & 2.937    \\
         & 15 & 8.251  & 2.811  & 2.935    \\
         & 20 & 8.249  & 2.810  & 2.935    \\
         & 25 & 8.248  & 2.810  & 2.935
\end{tabular}
\label{table1}
\end{table}
%
%
\mediumtext
 \begin{table}
  \caption{Square of the pseudoscalar bound-state mass $E^2$, and
           the lowest eigenvalue $\lambda^{\pm}$ of the matrix $H^{\pm}$
           as a function of the number of basis states for various
           values of the mass. The calculation were done
           assuming scalar confinement and an oscillator parameter
           of $\beta$=0.25 GeV.}
\begin{tabular}{ccccc}
   Mass [MeV]  & $\#$ states & $E^2$ $[\mbox{GeV}^2]$ & $\lambda^+$[GeV]
               &  $\lambda^-$[GeV] \\ \hline
    300  & 5  & -0.799   & 1.925  & -0.188   \\
         & 10 & -11.898  & 1.912  & -1.739    \\
         & 15 & -26.945  & 1.911  & -3.096    \\
         & 20 & -44.022  & 1.911  & -4.292    \\
         & 25 & -62.387  & 1.911  & -5.371    \\ \hline
    500  & 5  &   2.864  & 2.195  &  1.027    \\
         & 10 &  -0.916  & 2.177  & -0.137   \\
         & 15 & -11.240  & 2.175  & -1.301    \\
         & 20 & -24.254  & 2.175  & -2.372    \\
         & 25 & -39.003  & 2.175  & -3.359    \\ \hline
    900  & 5  &  6.821   & 2.844  &  2.362      \\
         & 10 &  6.752   & 2.814  &  2.231      \\
         & 15 &  6.752   & 2.811  &  1.513      \\
         & 20 &  6.751   & 2.810  &  0.733      \\
         & 25 & -0.465   & 2.810  & -0.041     \\
\end{tabular}
\label{table2}
\end{table}
%
%
%

\vfill
\eject
\end{document}